\documentclass[aps,prl,twocolumn]{revtex4}
\usepackage{graphicx,amssymb,amsmath}

\newcommand{\be}{\begin{equation}}
\newcommand{\ee}{\end{equation}}
\newcommand{\bea}{\begin{eqnarray}}
\newcommand{\eea}{\end{eqnarray}}
\newcommand{\LL} {\langle\!\langle}
\newcommand{\RR} {\rangle\!\rangle}
\newcommand{\la}{\langle}
\newcommand{\ra}{\rangle}
\newcommand{\lp}{\left(}
\newcommand{\rp}{\right)}

\begin{document}

\title{Measurement of Counting Statistics of Electron Transport in a Tunnel Junction}
\author{Yu. Bomze$^a$,  G. Gershon$^a$, D. Shovkun$^{a,b}$,
L.\,S. Levitov$^c$, and
M. Reznikov$^a$}
\affiliation{$^a$ Department of Physics and Solid State Institute,
Technion-IIT, Haifa 32000, Israel\\
$^b$ Institute for Solid State Physics, Chernogolovka, Russia\\
$^c$ Department of Physics,
Massachusetts Institute of Technology, 77 Massachusetts
Ave, Cambridge, MA 02139}

\begin{abstract}
We present measurements of the time-dependent fluctuations in
electrical current in a voltage-biased tunnel junction. We were able
to simultaneously extract the first three moments of the tunnel
current counting statistics. Detailed comparison of the second and
the third moment reveals that counting statistics is accurately
described by the Poissonian distribution expected for spontaneous
current fluctuations due to electron charge discreteness, realized
in tunneling transport at negligible coupling to environment.
\end{abstract}
\maketitle

Photon counting statistics~\cite{MandelWolf,WallsMilburn} is
a key technique
of quantum optics, which is routinely used to study
temporal and spectral distribution of electromagnetic field, and
to characterize the complexity of optical states,
such as photon coherence, entanglement, and squeezing.
In contrast, the electron counting and noise statistics, which are expected to
provide new insights into quantum transport,
is essentially at infancy as experimental subject,
with the first advances made only
recently~\cite{Reulet1,Rimberg,Fujisawa,Delsing,ReuletPRL}.

Electron counting proves
to be far more challenging than
that of photons primarily because of
the extremely high frequency of
electron passage events at typical current intensity,
which requires fast charge detection.
While in some cases Coulomb blockade in
quantum dots can be used to localize electrons and suppress tunneling,
bringing the tunneling
frequency down to the radio frequency range~\cite{Rimberg,Fujisawa,Delsing},
the time resolution needed to measure free, nonlocalized
electrons remains a challenge.

Another difficulty stems from the simple fact that,
while photons do not interact, the electrons do.
The electric
field fluctuations produced by the electrons which are being measured
can propagate out to other parts of the circuit (`environment')
and perturb it. In turn,
the noise due to the environment, modulated by the signal, can
couple back to the region of interest,
strongly affecting the measured
signal~\cite{KNB02,Nagaev02,Beenakker03,Reulet1,ReuletPRL,Geneva_group,Gutman05}.

The electron problem is especially rich and
intriguing:
unlike photons,
the current fluctuations are measured
without extracting electrons out of the system.
The electrons remain part of
the many-body system while
being detected, allowing
quantum phenomena to fully manifest themselves in electric noise.
This leads to a number of dramatic effects observed in electric noise,
such as, notably, elementary charge transmutation in
the Fractional Quantum Hall effect~\cite{depicciotto97,glattli97,reznikov99}
and
charge doubling in the Andreev scattering regime
in NS junctions~\cite{Kozhevnikov2000}.

The regime in which electric fluctuations occur spontaneously due to
charge discreteness, rather than due to thermal fluctuations, is
realized at sufficiently low temperatures~\cite{lesovik89,khlus87,BlanterReview}. It
was first demonstrated about 10 years ago in semiconductor point
contacts~\cite{reznikov95,kumar96} and in mesoscopic
wires~\cite{martinis96,schoelkopf97}. Current fluctuations in this
regime are traditionally analyzed using the noise frequency
spectrum. However, a more detailed
description\cite{levitov93,Nazarov03} is provided by the statistics of the
charge $q(\tau)$ transmitted through a conductor during a fixed time
interval $\tau$ which, in principle, can be small or large compared
to the time $\tau_0=e/I$. The low frequency shot noise is just the
second central moment of the transmitted charge distribution $P(q)$
at long times $\tau\gg\tau_0$:
\be
S^{(2)}=\la I_{\omega}^\ast I_{\omega}\ra_{\omega=0}=\frac{\la\Delta q^2(\tau)\ra}{\tau}=
\frac{\la (q(\tau)-\bar q)^2\ra}{\tau}
\ee
with $\la ...\ra$ a shorthand for $\int ...P(q,\tau)dq$.
The  counting
distribution $P(q)$, which encodes the full
information about noise statistics, was studied in various regimes,
including mesoscopic scattering~\cite{levitov93}, photon-assisted transport~\cite{ivanov93},
and transport in NS junctions~\cite{muzykantskii,nazarovNS},
where charge doubling takes place.
Specifics of the tunneling regime were considered in~\cite{LR01,sukhorukov}.

The present work
extends the shot noise measurements beyond the
second moment in a way uncorrupted by the presence of
electromagnetic environment.
The results agree with the expectation for Poisson
process, thus paving the way to investigation
of counting statistics.
We measure current fluctuations in a
tunnel junction by detecting the probability distribution of voltage
across a load resistor, $P(V)$. The latter
is directly related to
$P(q)$ when the load resistance is made much smaller
than the tunnel resistance.
The knowledge of $P(q)$ allows, in principle, to obtain
all moments of $q$.
However, with the measurement times $\tau\gg\tau_0$,
due to the central limit theorem,
the high moments become increasingly dominated by the lower order moments.
This makes the \emph{irreducible} parts of the moments, the cumulants,
which contain new information, increasingly
difficult to extract. Thus here we shall focus on the third central moment
$\langle\Delta q^3\rangle = \la (q-\bar q)^3\ra$
which coincides with the third cumulant $\LL q^3\RR$.

The expected
statistics is different in the
voltage-biased and current-biased regimes. The former case
is described by the
rate of the attempts to pass equal to $eV/h$ and a binomial
distribution of successes~\cite{levitov93}. At weak
transmission,
realized in tunnel junction, the
distribution assumes Poisson form,
with the low frequency spectral density $S^{(k)}$ of the
$k$-th cumulant, corresponding to the `long' measurement time $\tau\gg\tau_0$,
expressed as
\be \label{Sk}
S^{(k)}=\frac{\LL q^k\RR}{\tau}=e^{k-1}I
\ee
In contrast,
in a current-biased sample,
the rate of
successes is fixed
at $I/e$, and the
attempts to pass,
described by fluctuating time-dependent voltage
across the sample, are characterized by Pascal
distribution~\cite{KNB02}. A general cascade approach to high-order
noise statistics in the diffusive regime was developed in
Ref.~\cite{Nagaev02}, 
the role of external circuit was considered in
Ref.~\cite{Beenakker03} (see also \cite{Geneva_group}).

In the first measurement of the third moment $S^{(3)}$ 
Reulet~\emph{et al.}~\cite{Reulet1,ReuletPRL} 
used a low impedance tunnel junction
($R_{s} \simeq 50$\,Ohm) as a noise source, with a parallel
$50$\,Ohm load impedance $R_{l}$ of the cable used to feed the
voltage to an external amplifier. The initial results~\cite{Reulet1}
distinct difference from the Poissonian, in both magnitude and sign,
suggested that $S^{(3)}$ is dominated by the effects
extraneous to the junction.
A theoretical investigation~\cite{KNB02,Beenakker03} followed
which
clarified the importance of the
external circuit for correct interpretation of the
results, obtaining the third moment of the form
\be \label{env}
S^{(3)}=S_s^{(3)}- 3R_{s,l}
(S_s^{(2)}+S_l^{(2)})\frac{\partial(S_s^{(2)}-S_l^{(2)})}{\partial
V}
\ee
with $R_{s,l}=R_s R_l/(R_s+R_l)$ the impedance of the sample in
parallel with the load, $S_s^{(2)}$ and $S_l^{(2)}$ the sample
and load noise,
and $S_s^{(3)}$
generated by the voltage-biased sample. The load resistor, being
macroscopic, is not expected to produce a third
cumulant~\cite{Nagaev02,BlanterReview}.
The measurement~\cite{Reulet1, ReuletPRL},
which due to $R_l\simeq R_s$ was neither fully in the voltage-, nor
     in the current-biased regime,
was dominated by the
voltage-dependent $S_s^{(2)}$ in the last term of Eq.(\ref{env}).
Only at room temperature, due to low $\partial S_s^{(2)}/\partial
V$, the result had the sign of $S_s^{(3)}$.

\begin{figure} [tbp]
 \centerline{ \includegraphics [width=2.8in]
 {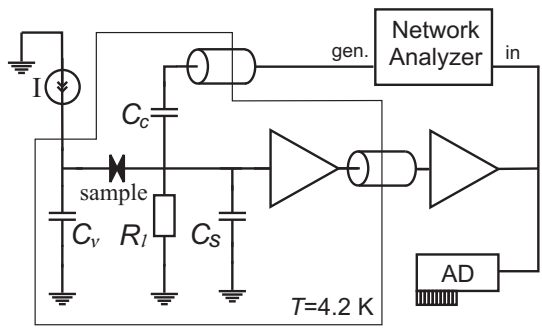} }
 \caption{Schematic of the experimental setup. The current source
 drives constant {\it average} current $I$ while the capacitor $C_v$
fixes the voltage across the sample and the load at relevant
frequencies. The bandwidth is determined by the load resistance
$R_{l}$ and stray capacitance $C_s$. The voltage fluctuations across
the load resistor are amplified and digitized by a 12 bit
analog-to-digital convertor (A/D) obtained from Ultraview  Corp.
with 5\,ns sampling time and 20\,ns interval between the samples.
The amplifier input impedance is determined by $R_{l}$. }
 \label{Setup}
\end{figure}

To measure the {\it intrinsic} $S^{(3)}$
free of the admixture of the second moment, we
use a new method suggested
and analyzed in Ref.\cite{LR01} (Fig.\ref{Setup}). Current
fluctuations generated by the sample (voltage-biased tunnel junction
of high impedance) produce voltage fluctuations on the load resistor
$R_l$ which are amplified and analyzed with computer.
The statistics of voltage fluctuations on $R_l$ is identical
to that of the intrinsic current fluctuations in the junction, provided
that $R_l$ is much smaller than the junction differential resistance.

The main source of errors in the measurement of the $k$-th cumulant
$\LL q^k \RR$ of the distribution $P(q)$ is statistical. In order to
estimate the signal-to-noise ratio the measured value $\LL q^k \RR$
should be compared to its variance
${\rm var}(\Delta  q^k)$ due to both sample and amplifier noise. The
variance is expressed through the central moments of the order $2k$.
The variance of an \emph{even} order for a generic distribution can
be estimated with the help of the central limit theorem, using
Gaussian statistics:
\be \label{eq:deltak}
 {\rm var}(\Delta  q^k)= \lp\langle \Delta
q^{2k}\rangle\rp^{1/2}\simeq \lp (2k-1)!!\rp ^{1/2} \la \Delta
q^2\ra ^{k/2}
\ee
Here
the variance $\la \Delta q^2\ra $
in Eq.(\ref{eq:deltak}) is due to
the sample, amplifier
and load noise, $S^{(2)}=S^{(2)}_s+S^{(2)}_l+S^{(2)}_a$, assuming all
three to be uncorrelated.
The signal-to-noise ratio for a single measurement of the third
cumulant is thus estimated as a ratio of $S_3\tau$ to $15^{1/2}
(S_2\tau)^{3/2}$, and  therefore it is beneficial to reduce the
sampling time $\tau$ to improve sensitivity,
in accord with the central limit theorem.
However, the amplified signal is correlated at
short times due to finite bandwidth of the input circuit. This makes
the effective sampling time restricted from below by the sample parasitic
capacitance, $C_s$, of both intrinsic and stray kind,
and by the effective output resistance of the sample in parallel
with the load resistance of the amplifier, $R_l$, so that $\tau_{\rm
eff} \simeq R_lC_s$. Repeating the measurement $N$ times would
further reduce fluctuations by a factor $\sqrt{N}$, assuming
statistically independent successive measurements. However, since
the measurements separated in time by less than $\tau_{\rm eff}$ are
correlated, the maximal improvement is
of the order of
$\sqrt{{\cal
T}/\tau_{\rm eff}}$, where $\cal T$ is the measurement time. In the
case of a high impedance tunnel
junction, the noise is dominated by the resistor $R_l$ thermal noise,
$2k_BT/R_l$. The measurements were performed at 4.2~K
to reduce both this
noise and the noise of the amplifier. Ignoring for now both the shot noise
produced by the sample and the amplifier noise, we estimate the
signal-to-noise ratio as
\be \label{eq:sn1} S/N=\frac{e^2 I \sqrt{\cal T}}{\sqrt{15}( 2k_B
T/R_l)^{3/2}\sqrt{\tau\tau_{\rm eff}}} \ee
Replacing $\tau$ by $\tau_{\rm eff}$ and plugging in
Eq.(\ref{eq:sn1}) we find the signal-to-noise ratio $S/N\propto
R_l^{1/2}C_s^{-1}$. It is therefore clear that it pays to decrease
$C_s$ and increase $R_l$ to improve $S/N$. We placed a cold
amplifier in the vicinity of the sample to reduce as much as
possible the capacitance $C_s$. The choice of the $R_l$ is
restricted by the desire to keep most of the signal at frequencies
well above $1/f$ corner of the amplifier. We chose $R_l=9.1\,{\rm
kOhm}$ which together with the stray capacitance of about $4\,{\rm
pF}$ gives $\tau_{\rm eff} = R_l C_s \simeq 40\,{\rm ns}$ and signal
bandwidth $(2\pi \tau_{\rm eff})^{-1}\approx 4\,{\rm MHz}$. To keep
the sample under voltage bias, we introduced a capacitor $C_v$ which
kept the voltage constant across the sample-load circuit at all
relevant frequencies.

We used tunneling through a $30 {\rm {\AA}}$ thick  ${\rm SiO_2}$
gate oxide of a $p$-channel Si field-effect transistor to produce a
shot noise. In this system tunneling occurs only under negative gate
voltage required to induce a hole channel. The differential
resistance of the barrier $R_s=(\partial I/\partial V)^{-1}>10^7\
{\rm Ohm}$ was much bigger then $R_l$, placing the sample securely
in the voltage-biased regime.
Indeed, the maximal
contribution of the second term in Eq.(\ref{env}),
\be
-3 Z (eI+2k_B T/R_l)e\frac{\partial I}{\partial V}\approx -\frac{6ek_B T}
{(R_l+R_s)},
\ee
is estimated as $5.2\times 10^{-48}\ {\rm A^3/Hz^2}$,
which is two orders of magnitude
smaller then $e^2I$ (Fig.~\ref{S3}).

\begin{figure}[tbp]
 \centerline{ \includegraphics [width=3.2in] {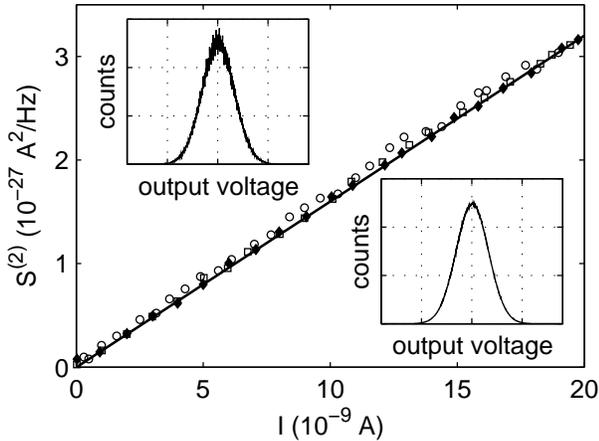} }
 \vspace*{-3mm}
 \caption{Measured value of the noise, $S^{(2)}$,
as a function of the current.
Different markers represent the data of three different measurements.
The straight line $S^{(2)}=eI$,
corresponding to the expected noise intensity~\cite{factor2}, is shown for guidance.
Upper
inset: raw voltage histogram sample; Lower inset:
cleaned histogram, illustrating the result of normalizing
with the A/D converter calibration (see text).
}
 \label{S2}
\end{figure}
\begin{figure} [tbp]
 \centerline{ \includegraphics [width=3.4in] {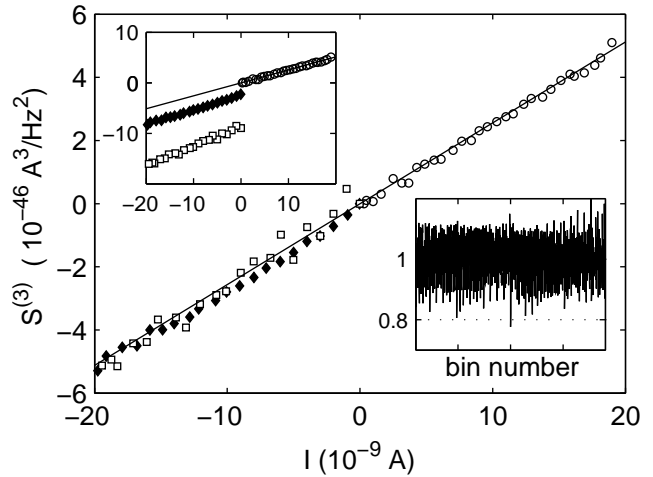}} 
 \vspace*{-3mm}
 \caption{
Measured third cumulant $S^{(3)}$ of the transmitted charge, obtained
separately for different current directions. (Markers are the same
as in Fig.\,\ref{S2}, the straight line is $S^{(3)}=e^2I$.) Upper
inset: $S^{(3)}$ {\it vs.} $I$ without amplifier nonlinearity
correction; Lower inset: normalized histogram of the linearly swept
signal, used to calibrate the A/D converter (see text). } \label{S3}
\end{figure}

We record
the probability distribution function $P(V_a)$
of the amplified voltage $V_a$ (Fig.\,\ref{S2}, upper inset).
To clean the obtained histogram we correct for
nonequal bin widths of the A/D converter by calibrating the latter
against linearly swept signal (Fig.\,\ref{S3}, lower inset).
The effect of normalizing $P(V_a)$ with
the A/D converter calibration, illustrated in Fig.\,\ref{S2} lower inset,
is two-fold. First, the histogram
loses the erratic features (`grass') on the smallest scale.
Simultaneously, the envelope is somewhat corrected
on large scale.
The latter effect, due to averaging, has
longer stability time, which is fortunate, since the
`grass' features, while less stable, were found to
have little effect on the second and the
third cumulant.

For a linear circuit,
the amplified voltage $V_a(\omega)$
and the input current $i(\omega)$ Fourier
components
are proportional:
$V_a(\omega)=Z(\omega)K(\omega)i(\omega)$,
with $Z$
the load impedance
and $K$ the amplification.
Therefore the
third cumulant
is related to the respective spectral densities of $V_a$ as
%
\be
\label{Va3}
\la V_a^{(3)}\ra = \frac{S^{(3)}}{(2\pi)^2} {\iint}_{-\infty}^{+\infty}
A(\omega)A(\omega')A(-\omega-\omega')
d\omega d\omega'
\ee
with $A(\omega)=Z(\omega)K(\omega)$.
For the broadband amplifiers used, the amplification $K(\omega)$ was
almost frequency independent between the low $\sim 0.5\ {\rm MHz}$
and high $\sim 25 \ {\rm MHz}$ frequency cutoffs intentionally
introduced to filter the $1/f$ amplifier noise and the wide-band
noise at frequencies well above $(2\pi \tau_{\rm eff})^{-1}$.
The complex product
$Z(\omega)K(\omega)$
was obtained from calibration,
introducing a known current through a small
capacitor $C_c$ typically of 2.4~pF. We then used
Eq.(\ref{Va3}) for $V_a^{(3)}$ and a similar expression for $V_a^{(2)}$
to extract $S^{(3)}$ and $S^{(2)}$.
The second cumulant $S^{(2)}$ combines
the shot, thermal and amplifier noise contributions. Since only the
former depends on current, the amplifier noise and the thermal noise,
obtained from $S^{(2)}$ at zero current, can be easily
subtracted~\cite{CurrNoise}.
The resulting noise $S^{(2)}$, shown in Fig.\,\ref{S2},
varies linearly with current  as
expected ($eV\simeq 4\,{\rm V}\gg k_BT$ at all currents).
The measured value agrees with the expected value, $eI$, within
calibration error of 5\%.

In order to extract properly the third cumulant of the voltage, one
should account for the effect of amplification nonlinearity
which can mix $S^{(2)}$ with $S^{(3)}$. Let us consider the
amplifier nonlinearity, $V_{out}=K(V_{in}+V_{in}^2/U)$, with $1/U$
the nonlinearity parameter, which which contributes to the cumulants
$\la \delta V_{out}^2\ra$ and $\la \delta V_{out}^3\ra$
as follows:
\be \label{nonlin2} \la \delta V_{out}^2\ra/K^2-\la \delta V_{in}^2
\ra =2 \la \delta V_{in}^3\ra  /U + 3 \la \delta V_{in}^2\ra ^2 /U^2
\ee
\be \label{nonlin3} \la \delta V_{out}^3\ra/K^3-\la \delta V_{in}^3
\ra =9 \la \delta V_{in}^2\ra ^2 /U\ee
The right hand side of the Eq.(\ref{nonlin3}) would
affect the value $S^{(3)}$ calculated from $\la \delta
V_{out}^3\ra$.
The parameter $U$ can be
estimated by applying an ac signal and measuring its
second harmonic. We took special care to reduce nonlinearity of the
amplifier track, especially the last amplifier in the chain.
We managed to make it comparable to the nonlinearity
of the cryogenic amplifier
which was different for different measurements,
depending on the regime. We found
$U$ to exceed $20\,{\rm mV}$, and thus having negligible
effect on $S^{(2)}$.
However, estimates show that
the nonlinearity correction
to $S^{(3)}$
can be as large as $10^{-45}\,{\rm A^3/Hz^2}$.
We attribute the third cumulant at zero current (see the
upper inset of Fig.\,\ref{S3}) to this nonlinearity and determine
$U$ as the ratio $9\la \delta V_{in}^2\ra^2_{I=0}/\la \delta
V_{out}^3\ra_{I=0}$. Using this $U$ we subtract the nonlinearity
contribution, Eq.(\ref{nonlin3}), from the data at all currents.
This procedure, as illustrated in Fig.\,\ref{S3} upper inset,
restores zero value at $I=0$, but has little effect on the slope of
the dependence $S^{(3)}$ \emph{vs.} $I$. We therefore conclude that
the nonlinearity, while necessary to account for to improve
accuracy, is not strong enough to compromise the measurement of
$\partial S^{(3)}/\partial I$. The results, shown in Fig.\,\ref{S3},
are found to be in excellent agreement with
the Poissonian third cumulant $e^2I$.

As an additional check, we also reversed the current direction.
Since we had to apply negative voltage on the gate to
induce holes, it required rebonding the sample (see Fig.\,\ref{S3},
upper inset). We found the third
cumulant to be an odd function of current, as expected.


 
\emph{Note added:}
Finally, we note that 
the literature is not entirely unanimous regarding the
Poissonian character of tunneling transport. Dissent is
exemplified by Ref.~\cite{Lesovik03}, 
predicting a new "quantum regime" at sufficiently
low frequencies, limited by inverse flight time in a noninteracting
fermion model. Ref.~\cite{Lesovik03} obtains $S^{(3)}$ for the tunneling current of sign
opposite to Poissonian $S^{(3)}$ and also of much smaller magnitude, 
quadratic in transmission rather than linear as in Eq.(\ref{Sk}) above. 
The conditions stated
in Ref.~\cite{Lesovik03} are fulfilled in our experiment: the time interval between
electron tunneling and its detection, estimated from EM signal propagation 
speed, 
is of order $3\cdot 10^{-11}\,{\rm s}$, which is much shorter than the
sampling time, $5\,{\rm ns}$, 
placing the measurement securely at low frequency in the sense of
Ref.~\cite{Lesovik03}. The results \cite{Lesovik03} are thus
not in accordance with our observations.

In summary, we present the first measurements of the
charge counting
statistics in \emph{voltage-biased} tunnel junction up to the third
cumulant. The results, obtained by analyzing the distribution of transmitted
charge, are in excellent agreement with expectations
for Poissonian process, making electron counting statistics
amenable to experimental investigation.

We are indebted to O.\,Prus for his contribution at the early
stage of the work. 
One of us (M.R.) is indebted to G.\,Lesovik for attracting his attention 
to the problem.
This research was supported by the US-Israel
Binational Science Foundation and (D.S. and Yu. B.)
by the Lady Davis Fellowship.


\begin{thebibliography}{qq}
\bibitem{factor2}
Our definition of noise moments differs by a factor of $2$ from
the one based on the positive frequency
representation for the noise power spectrum.
It seems more natural to use
the entire range of frequencies from $-\infty$ to $\infty$, which
brings the Schottky relation to the form $S_2=eI$.
\bibitem{CurrNoise} Since the sample differential resistance is much
bigger then the load $R_l$, the resistance seen by the amplifier
is current-independent and
thus the amplifier current noise contribution is negligible.
\end{thebibliography}
\end{document}